\documentclass{aa}  

\usepackage{graphicx}
\usepackage{txfonts}
\begin{document}

   \title{Blazar nature of high--$z$ radio--loud quasars}

   %\subtitle{I. Overviewing the $\kappa$-mechanism}

   \author{T. Sbarrato\inst{1}
            \and G. Ghisellini\inst{1}
            \and G. Tagliaferri\inst{1}
            \and F. Tavecchio\inst{1}
            \and G. Ghirlanda\inst{1}
            \and L. Costamante\inst{2}
          	}

   \institute{INAF -- Osservatorio Astronomico di Brera,
              	via Emilio Bianchi 46, I--23807 Merate, Italy\\
             	\email{tullia.sbarrato@inaf.it}
        \and
        	ASI -- Agenzia Spaziale Italiana, Via del Politecnico snc, I--00133 Roma, Italy}

   \date{}%Received September 15, 1996; accepted March 16, 1997}

% \abstract{}{}{}{}{} 
% 5 {} token are mandatory
 
  \abstract{
  We report on the {\it Swift}/XRT observation and classification of eleven blazar candidates at $z>4$. 
  These sources were selected as part of a sample of extremely radio--loud quasars, 
  in order to focus on quasars with jets oriented roughly close to our line--of--sight. 
  Deriving their viewing angles and their jets bulk Lorentz factors was crucial for a strict 
  blazar classification, and it was possible only thanks to X--ray observations.
  Out of eleven sources, five show strong and hard X--ray fluxes, that allow their blazar classification, 
  two are uncertain, three host relativistic jets that we observe just outside their beaming cone (i.e.\ 
  are not strictly blazars), while one went undetected by {\it Swift}/XRT.
  Following this approach, we were able to trace the $>10^9M_\odot$ active supermassive black hole population 
  hosted in jetted active galactic nuclei.
  At $z\geq4$ the massive jetted sources are likely predominant in the overall quasar population:
  this calls for a deep review of our understanding of the first supermassive black holes formation and evolution. 
  Jets are indeed key actors in fast accretion, and must be searched for across the whole high redshift quasar population.
  A note of caution must be added: radio--loudness and in general radio features at high redshifts seem not to perfectly 
  reflect high--energy properties. A strong effect due to interaction with CMB radiation is surely in place, 
  that quenches the radio emission with respect to the X--rays, but also 
  more frequent occasions for the jet to be bent seem to play a relevant role in this matter.
  Classifications and population studies thus must be carefully performed, in order not to be confused by these inconsistencies. 
  }
%   {Optional context section
%   }
%  % aims heading (mandatory)
%   {Mandatory aims section.
%   }
%  % methods heading (mandatory)
%   {Mandatory methods heading.
%   }
%  % results heading (mandatory)
%   {Mandatory results section.
%   }
%  % conclusions heading (optional), leave it empty if necessary 
%   {}


   \maketitle
%
%-------------------------------------------------------------------

\section{Introduction}

The active supermassive black hole census is becoming very populated.
At extreme redshifts ($z>5.7$) $\sim300$ quasars are currently known \citep{fan99,banados16,jiang16}, 
while few dozens of radio--loud Active Galactic Nuclei are known at $z>4.5$, with only few 
at redshift larger than 6, being much rarer than their radio--quiet or radio--silent counterparts. 
The fraction of radio--loud sources among the total high--$z$ quasar population is currently consistent with the local Universe \citep{banados15}, i.e.\ $\sim10\%$ of known quasars have a radio flux more than 10 times brighter than its emission in the blue band.
Radio--loudness is generally associated to the presence of strong relativistic jets, since extended lobes and/or a relativistically beamed structure are the sole responsible for such strong radio emissions.

This distant active supermassive black hole population  
is currently mainly composed by extremely massive, highly accreting objects. 
All  known high--redshift quasars have black hole masses $M>3\times10^8M_\odot$ 
and accretion rates $>10\%$ of the Eddington limit, 
with the large majority more massive than $10^9M_\odot$ and around $30\%L_{\rm Edd}$.
This is clearly an observing bias, but the presence of these sources 
have raised many issues on the formation and evolution of the 
first supermassive black holes. 
\cite{volonteri12} and \cite{inayoshi20} reviewed the possible mechanisms to build up $M>10^9M_\odot$ in a short amount of time ($<1$ Gyr). 
In order to assemble such massive black holes, a $M>10^4M_\odot$ black hole seed is necessary if the accretion occurs in an Eddington--limited regime. 
The formation of this kind of seed is not straightforward: collapse of a massive gas cloud might be prevented by fragmentation due to cooling and star formation activity.
Clearly, the accretion might be faster if super--Eddington phases are encountered during the SMBH evolution, introducing the known difficulties of sustaining a continuous super--critical regime. 
Nevertheless, a super--Eddington accretion or phases would allow a smaller black hole seed, avoiding the issue of fragmentation during its formation. 
Relativistic jets have been proposed as important tools to allow the fast accretion of early quasars \citep{jolley08,ghisellini13,regan19}, and in fact their distribution in the early Universe is crucial to understand their link to the formation of the first supermassive black holes. 

Tracing the presence of jetted sources at $z>4$ is not an easy task: the radio surveys start reaching their sensitivity limit in detecting distant sources. 
The effect of beaming on emitted radiation becomes thus crucial in order to make jets visible at high redshift. 
Blazars have thus a peculiar advantage in their selection and observation. 
Jetted AGN are classified as blazars when their jets are aligned to our line--of--sight, allowing their jets to dominate the overall emission across the whole electromagnetic spectrum. 
Their spectral energy distribution (SED) is characterized by two broad humps, produced by synchrotron and inverse Compton (IC) emission (at low and high frequency, respectively). 
The synchrotron radiation is visible down to the MHz--GHz frequency range, while IC normally reaches $\gamma$--ray energy range, in some cases up to the TeV domain.
The latter is normally the smoking gun of blazars emission: synchrotron is in fact responsible for strong radio emission in misaligned jetted AGN as well, even if less intense or with different spectral shapes.
For this reason, in the local Universe and up to $z\sim3-3.5$ blazars are easily detected and classified thanks to all--sky high--energy facilities, such as {\it Fermi}/LAT.
Because of its sensitivity limit and the redshift of the high--energy component at $z>3.5$, at such distances blazars are not systematically detectable. 
A more focussed approach is thus needed. 
The solution lies in pointed X--ray observations: 
instead of performing an all--sky unbiased search, a sample of blazar candidates must be collected in order to study the first section of their IC emission, if present. 
We selected reliable candidates on the basis of their optical features and strong radio brightness (see Section \ref{sec:sample}). 

In the following, we present X--ray observations of eleven sources from \cite{sbarrato13a} and their broad--band SED modeling. 
The sample selection and data collection and analysis are shown in Section \ref{sec:sample}. 
Section \ref{sec:model} details the model we used to interpret the broad--band Spectral Energy Distribution (SED) and derive the key physical parameters of our sample. 
In Section \ref{sec:double_jet} we discuss how radio and X--ray features in a few cases point towards different conclusions on jet inclination angle. 
Section \ref{sec:early_jet} will finally discuss the implications on the number density of jetted quasars given by our findings, and how it impacts our knowledge of early formation and evolution of supermassive black holes.
Finally, Section \ref{sec:conclusion} summarizes our work and findings. 
In this work, we adopt a flat cosmology with 
$H_0 = 70$ km s$^{-1}$ Mpc$^{-1}$ and $\Omega_{\rm M}=0.3$.

   %Introducing the ultramassive black holes in the early Universe, 
   %their problematic but necessary fast formation and evolution and 
   %the current observational situation.

%-------------------------------------------------------------
\begin{table*} 
\centering
\small
\begin{tabular}{lccccccc}
\hline
\hline
\vspace{0.1cm}
 Coord.   &$z$ &$R$ &ObsID & Exposure  &$F^{\rm obs}_{\rm 0.3-10 keV}$  &$\Gamma_{\rm X}$   &$N_{\rm H}$  \\
 & &  & & (ks)  &(${\rm erg/cm^2/s}$)  &   &(${\rm10^{20} cm^{-2}}$) \\
~[1]      &[2] &[3]    &[4]             &[5]    &[6]    &[7]    &[8]\\
\hline   
030437.21+004653.5  &4.305 &2410    &85420001   & 24.4  &$<3.75\times10^{-14}$ &   ---   &$6.4$ \\
\vspace{0.1cm}
 & & &85420010 & & & & \\
081333.32+350810.8  &4.922 &610     &85966012    & 37.6  &$3.69^{+1.24}_{-1.17}\times10^{-14}$    &$1.80^{+0.56}_{-0.54}$       &$4.9$\\ 
 & & &85966017-018 & & & & \\
\vspace{0.1cm}
 & & &85966020 & & & & \\
085111.59+142337.7  &4.307 &270     &87238013-014    & 38.0  &$2.93^{+0.39}_{-0.34}\times10^{-13}$   &$1.34\pm0.20$   &$3.2$\\ 
 & & &87238016-018 & & & & \\
\vspace{0.1cm}
 & & &87238020 & & & & \\
103717.72+182303.0  &4.051 &214     &87239003    & 37.6  &$3.27^{+2.13}_{-1.07}\times10^{-14}$   &$1.32\pm0.71$       &$2.0$\\ 
 & & &87239006 & & & & \\
\vspace{0.1cm}
 & & &87239019 & & & & \\
105320.42--001649.7 &4.304 &149     &87240003-005    & 30.5  &$5.78^{+2.99}_{-1.43}\times10^{-14}$   &$1.45\pm0.56$    &$3.8$\\ 
\vspace{0.1cm}
 & & &87240009 & & & & \\
123142.17+381658.9  &4.137 &264     &85967001-002    & 35.6  &$7.37^{+2.02}_{-1.59}\times10^{-14}$   &$1.37^{+0.42}_{-0.43}$       &$1.3$\\ 
 & & &85967004 & & & & \\
\vspace{0.1cm}
 & & &85967006-013 & & & & \\
\vspace{0.1cm}
123503.03--000331.7 &4.723 &1493    &87250011    & 42.5  &$4.33^{+1.55}_{-1.31}\times10^{-14}$   &$1.44^{+0.60}_{-0.63}$       &$1.9$\\ 
\vspace{0.1cm}
124230.58+542257.3  &4.730 &631     &85968001    & 28.1  &$1.44^{+2.88}_{-0.61}\times10^{-14}$  &$1.75^{+1.43}_{-1.11}$       &$1.4$\\ 
\vspace{0.1cm}
141209.96+062406.8  &4.467 &852     &85421006    & 26.5  &$5.5^{+2.9}_{-1.9}\times10^{-14}$      &$1.88\pm0.31$             &$2.1$\\ 
165913.23+210115.8  &4.784 &637     &87251013    & 42.0  &$6.27^{+1.94}_{-1.13}\times10^{-14}$   &$1.71^{+0.41}_{-0.40}$       &$5.3$\\ 
 & & &87251021-022 & & & & \\
\vspace{0.1cm}
 & & &87251024-025 & & & & \\
\vspace{0.1cm}
231448.71+020151.1  &4.110 &2388    &85422001-015    & 25.0  &$5.4^{+3.0}_{-1.9}\times10^{-14}$      &$1.76\pm0.31$             &$4.7$\\  
\hline 
\end{tabular}
\vskip 0.4 true cm
\caption{Data analysis details.
Col.\ [1] RA \& Dec (J2000);
Col.\ [2]: redshift;
Col.\ [3]: radio--loudness;
Col.\ [4]: {\it Swift} Observation ID (dashes means that all ObdIDs in the range are considered);
Col.\ [5]: total exposure in ks;
Col.\ [6]: observed X--ray flux in the $0.3-10$keV range;
Col.\ [7]: photon index;
Col.\ [8]: Galactic absorption column.
}
\label{tab:data}
\end{table*}
%-------------------------------------------------------------

\section{Sample selection and data analysis}   
\label{sec:sample}

In \cite{sbarrato13a} we selected a sample of 19 spectroscopic blazar candidates,  
together with other 12 outside of the SDSS+FIRST spectroscopic footprint.
We followed up the first 4 of them with X--ray observations, and managed to classify them as blazars in \cite{sbarrato15}. 
We then obtained {\it Swift}/XRT observations for other eleven sources
which we will be presenting in this work. 

The sample selection criteria are based on optical and radio features. 
We started from the quasar catalog of the Sloan Digital Sky Survey \citep[SDSS,][]{york00} Seventh Data Release \citep[DR7,][]{schneider10}, selecting the highest redshift and most radio--loud sources. 
We focus on $z>4$ sources, and we only consider those with a radio--loudness $R=F_{\rm 5GHz}/F_{\rm 4400\AA}>100$ \citep[rest frame fluxes, ][]{kellermann89}. 
Powerful jets are generally associated with radio--loud AGN ($R>10$). 
When the jet is aligned close to our line--of--sight, the relativistic beaming is responsible for an over--boosting of the radio emission, leading to extreme $R$ values. 
Selecting $R>100$ allows us to focus only on the most likely aligned sources. 
This selection proved itself effective: the most radio bright sources out of the original sample were confirmed to be blazars
with accretion features consistent with Flat Spectrum Radio Quasars, i.e.\ fast accretors with strong emission lines, redder and brighter broad band SEDs \citep{sbarrato12,sbarrato13b,sbarrato15}.
We thus continue the identification process by observing the next probable blazar candidates 
within our sample.

\subsection{Swift/XRT observations and data}
\label{sec:data}

The analysis of the data from the X--Ray Telescope \citep[XRT;][]{burrows05} 
onboard the Swift satellite has been done by using
HEASOFT v 6.29 and the CALDB updated on 2022 January 1 and
by following standard procedures as described e.g.\ in \cite{sbarrato15}. 
Because of the low-statistics, X-ray data were analysed by
using unbinned likelihood \citep{cash79}.

One of the blazar candidates selected for this work (SDSS J030437.21+004653.5, $z=4.305$) is 
undetected by {\it Swift}/XRT after 24.4ks of observations. 
We report in Table \ref{tab:data} the $3\sigma$ upper limit in the $0.3-10$ keV flux. 
All other sources are well detected, with intense fluxes and more or less hard spectra. 
All results are reported in Table \ref{tab:data} and discussed in the following.

%-------------------------------------------------------------
%                                             Two column Table 
%-------------------------------------------------------------
%

\begin{table*} 
\centering
\tiny
\begin{tabular}{llllllllllllllllll}
\hline
\hline
 Coord.   &$z$ &$M$ &$L_{\rm d}$ &$R_{\rm diss}$ &$R_{\rm BLR}$ &$P^\prime_{\rm i}$  &$B$ &$\Gamma$ &$\theta_{\rm v}$  
&$\gamma_{\rm b}$ &$\gamma_{\rm max}$ &$s_1$  &$s_2$ &$P_{\rm j}$ \\
~[1]      &[2] &[3] &[4] &[5] &[6] &[7] &[8] &[9] &[10] &[11] &[12] &[13]  &[14] &[15] \\
\hline   
081333.32+350810.8  &4.922 &4e9   &6.2e46 (0.12) &1.2e3 (1e3)  &790 &5e42  &1.7 &13 &5.2 &150 &3e3 &0    &2.6 &3.8e46  \\
				  &4.922 &4e9   &6.2e46 (0.12) &1.2e3 (1e3)  &790 &5e43  &1.7 &13 &8   &150 &3e3 &0    &2.6 &1.6e47 \\
\\
085111.59+142337.7  &4.307 &4e9   &3.6e46 (0.07) &720   (600)  &603 &4e42  &1.1 &13 &3   &100 &4e3 &0.75 &2.7 &3.2e46 \\
				  &4.307 &4e9   &3.6e46 (0.07) &720   (600)  &603 &4e43  &1.3 &13 &6   &100 &4e3 &0.75 &2.7 &2.9e47 \\
\\
103717.72+182303.0  &4.051 &1e9   &2.6e46 (0.20) &180   (600)  &510 &2e42  &7.1 &13 &3.5 &250 &3e3 &0    &2.6 &1.2e46 \\
				  &4.051 &1e9   &2.6e46 (0.20) &180   (600)  &510 &2e43  &7.1 &13 &6   &250 &3e3 &0    &2.6 &5.1e46 \\
				  &4.051 &1e9   &2.6e46 (0.20) &180   (600)  &510 &8e43  &7.1 &13 &8   &250 &3e3 &0    &2.6 &1.8e47  \\
\\
105320.42--001649.7 &4.304 &1.5e9 &5.3e46 (0.27) &248   (550)  &725 &4e42  &3.7 &13 &5   &60  &4e3 &0    &2.6 &2.6e46 \\
				  &4.304 &1.5e9 &5.3e46 (0.27) &248   (550)  &725 &8e42  &3.7 &13 &6   &60  &4e3 &0    &2.6 &4.7e46 \\
				  &4.304 &1.5e9 &5.3e46 (0.27) &248   (550)  &725 &5e43  &3.7 &13 &8   &60  &4e3 &0    &2.6 &2.7e47 \\
\\
123142.17+381658.9  &4.137 &7e8   &1.8e46 (0.2)  &147   (700)  &427 &3e42   &4.4 &11 &3   &70  &3e3 &0    &2.6 &1.1e46  \\
				  &4.137 &7e8   &1.8e46 (0.2)  &147   (700)  &427 &3e43   &4.4 &11 &6   &70  &3e3 &0    &2.6 &9.7e46  \\
				  &4.137 &7e8   &1.8e46 (0.2)  &147   (700)  &427 &1e44   &5.5 &11 &8   &70  &3e3 &0    &2.6 &3.3e47  \\
\\
123503.03--000331.7 &4.723 &2e9   &1.8e46 (0.07) &660   (1100) &427 &1.5e42 &0.9 &13 &3   &50  &3e3 &0    &2.7 &9.5e45 \\
				  &4.723 &2e9   &1.8e46 (0.07) &660   (1100) &427 &1.5e43 &1.2 &13 &6   &50  &3e3 &0    &2.7 &7.6e46 \\
\\
124230.58+542257.3  &4.730 &3.5e9 &3.6e46 (0.08) &1050  (1000) &603 &6e42   &0.8 &13 &6   &300 &3e3 &1    &2.4 &4.7e46 \\
				  &4.730 &3.5e9 &3.6e46 (0.08) &1050  (1000) &603 &1e44   &1.0 &13 &10  &300 &3e3 &1    &2.4 &5.1e47 \\
\\
141209.96+062406.8  &4.467 &6e8   &4.9e46 (0.63) &216   (1200) &701 &3e42   &4.9 &11 &3   &80  &3e3 &1    &2.6 &2.5e46  \\
				  &4.467 &6e8   &4.9e46 (0.63) &216   (1200) &701 &7e42   &5.4 &10 &5   &100 &3e3 &1    &2.6 &4.1e46  \\
\\
165913.23+210115.8  &4.784 &2.5e9 &3.3e46 (0.1)  &450   (600)  &570 &3e42   &1.6 &13 &3   &300 &4e3 &1    &2.5 &1.6e46 \\
				  &4.784 &2.5e9 &3.3e46 (0.1)  &450   (600)  &570 &2e44   &1.6 &13 &8   &300 &4e3 &1    &2.5 &8.7e47 \\
\\
231448.71+020151.1  &4.110 &1.5e9 &2.9e46 (0.15) &225   (500)  &540 &5e42   &3.4 &10  &3  &200 &5e3 &-0.5   &2.8 &1.6e46  \\ 
				  &4.110 &1.5e9 &2.9e46 (0.15) &225   (500)  &540 &4e43   &3.4 &10  &6   &200 &5e3 &-0.5   &2.8 &6.3e46   \\ 
\hline 
\end{tabular}
\vskip 0.4 true cm
\caption{Input parameters used to model the SED.
Note that $R_{\rm BLR}$ is a derived quantity, not an independent input parameter.
It is listed for an easy comparison with $R_{\rm diss}$.
Col. [1] RA \& Dec (J2000);
Col. [2]: redshift;
Col. [3]: black hole mass in solar masses;
Col. [4]: accretion disk luminosity, in erg s$^{-1}$ and 
        (in parenthesis) in units of $L_{\rm Edd}$;
Col. [5]: dissipation radius in units of $10^{15}$ cm and (in parenthesis) in units of $R_{\rm S}$;
Col. [6]: size of the BLR in units of $10^{15}$ cm;
Col. [7]: power injected in the blob calculated in the comoving frame, in erg s$^{-1}$; 
Col. [8]: magnetic field in Gauss;
Col. [9]: bulk Lorentz factor at $R_{\rm diss}$;
Col. [10]: viewing angle in degrees;
Col. [11] and [12]:  break and maximum random Lorentz factors of the 
injected electrons;
Col. [13] and [14]: slopes of the injected electron distribution [$Q(\gamma)$] 
below and above $\gamma_{\rm b}$;
Col. [15]: total jet power calculated as the sum of all power components (in erg/s).
For all cases the X--ray corona luminosity $L_X=0.3 L_{\rm d}$.
Its spectral shape is assumed to be $\propto \nu^{-1} \exp(-h\nu/150~{\rm keV})$.
}
\label{tab:sed}
\end{table*}
%-------------------------------------------------------------

\section{The model}
\label{sec:model}

We model the broad band SEDs of our candidates including new 
X--ray observations, in order to interpret their nuclear and jet emission.
We will be able to classify them as blazars or not thanks to this model, 
that will allow us to estimate beaming factors and black hole masses and,  
therefore, to study the early jetted quasar population.

%\vskip 0.3 cm
%\noindent
\subsection{Black hole mass and accretion luminosity} 

Our blazars are very luminous, not only due to their beamed jet flux, but
also for their prominent accretion disk emission. 
Furthermore, the large redshift let us see the high frequency part of the
jet emission, close to its peak. 
And finally, being blazar candidate ensures that we are observing the jet at
a small viewing angle, that implies (most likely) that the disk is viewed face--on.

This means that we can use an accretion disk model to find out the two parameters 
of interest: the black hole mass $M$ and the bolometric
accretion luminosity $L_{\rm d}$, in a reliable way.
For simplicity, we use the standard \citet{shakura73} disk model,
assuming an efficiency $\eta$ (defined by $L_{\rm d}=\eta \dot M c^2$) 
fixed to the value of $\eta=0.083$.
For a given $\dot M$ (i.e. for a given $L_{\rm d}$),
the black hole mass regulates the peak frequency of the 
disk emission (heavier black holes have larger Schwarzschild radii, and
thus colder disks).
Consider also that, at the peak of the disk emission, $\nu L_\nu \sim L_{\rm d} /2$
\cite[see also][]{calderone13}.
If the peak of the disk spectrum is visible (as in the case of this sample), the uncertainties are
on average less than a factor 2, better than the virial method. 
Note that this approach in deriving black hole masses is particularly rewarding, given that at $z>4$ the virial method is mainly based on the CIV emission line, that introduces even larger uncertainties than the standard factor 3--4 because of the asymmetric profile that shows frequently.

For the accretion disk luminosity, our sources show no signs of contribution of
the jet emission in the optical--UV, making the estimate of $L_{\rm d}$ reliable. 
Note that, with respect to the parameter $L_{\rm bol}$ used by e.g. \citet{shen11},
there is a factor 2 difference (i.e. $L_{\rm bol} \sim 2 L_{\rm d}$) because 
$L_{\rm bol}$ includes the X--ray corona emission and the IR flux from the molecular torus.

A posteriori, we have checked that $L_{\rm d}$ is always below $0.3 L_{\rm Edd}$ 
(with one exception) and above 
$10^{-2} L_{\rm Edd}$, consistent with the use of a Shakura--Sunyaev disk model.

% ================================================

\subsection{The jet model}

%--------------------------------------------------
\begin{figure*} 
\vskip -2.5 cm
\hskip -0.25 cm
\includegraphics[width=9.4cm]{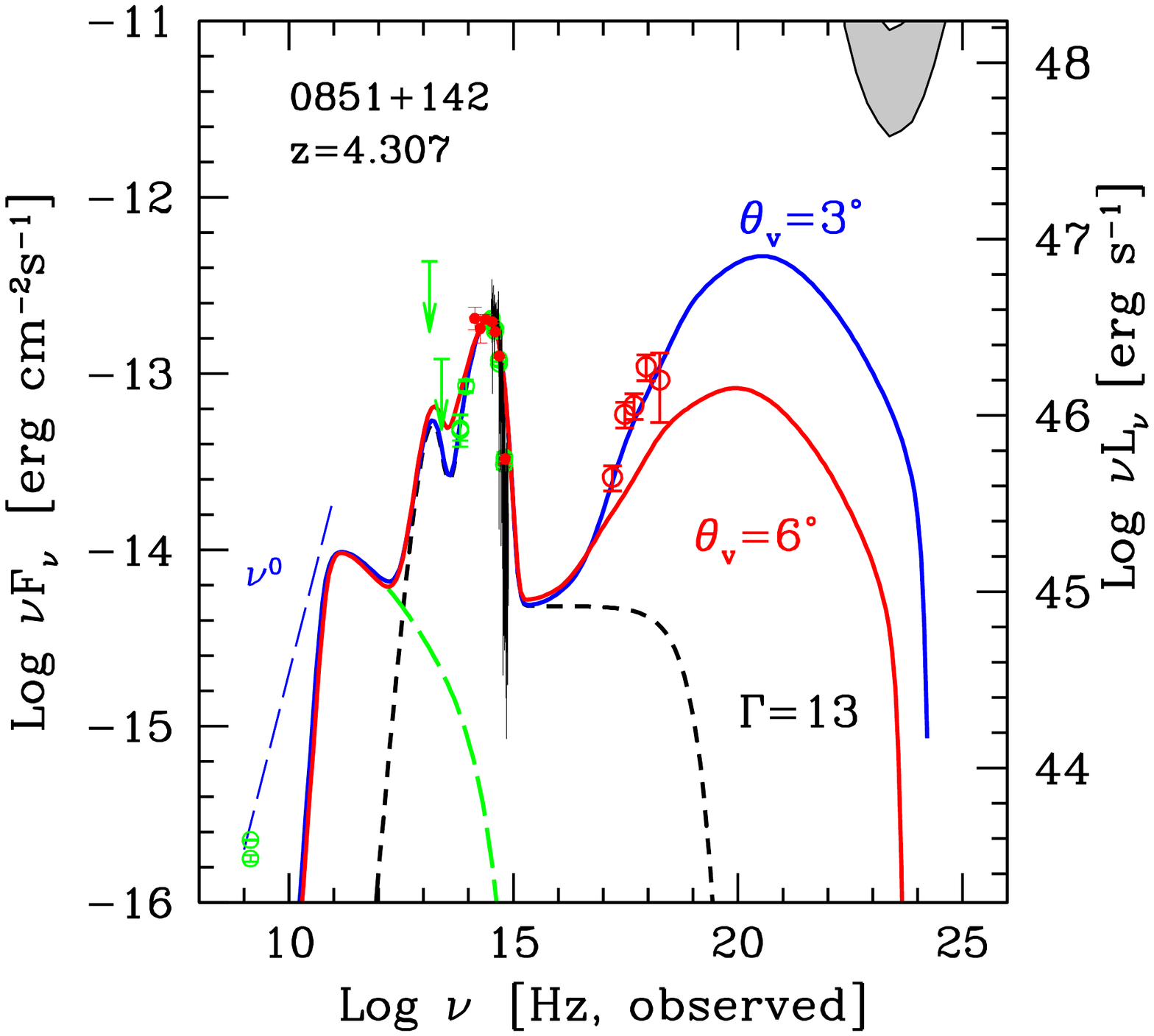} % ,width=12cm}
\hskip -0.25 cm
\includegraphics[width=9.4cm]{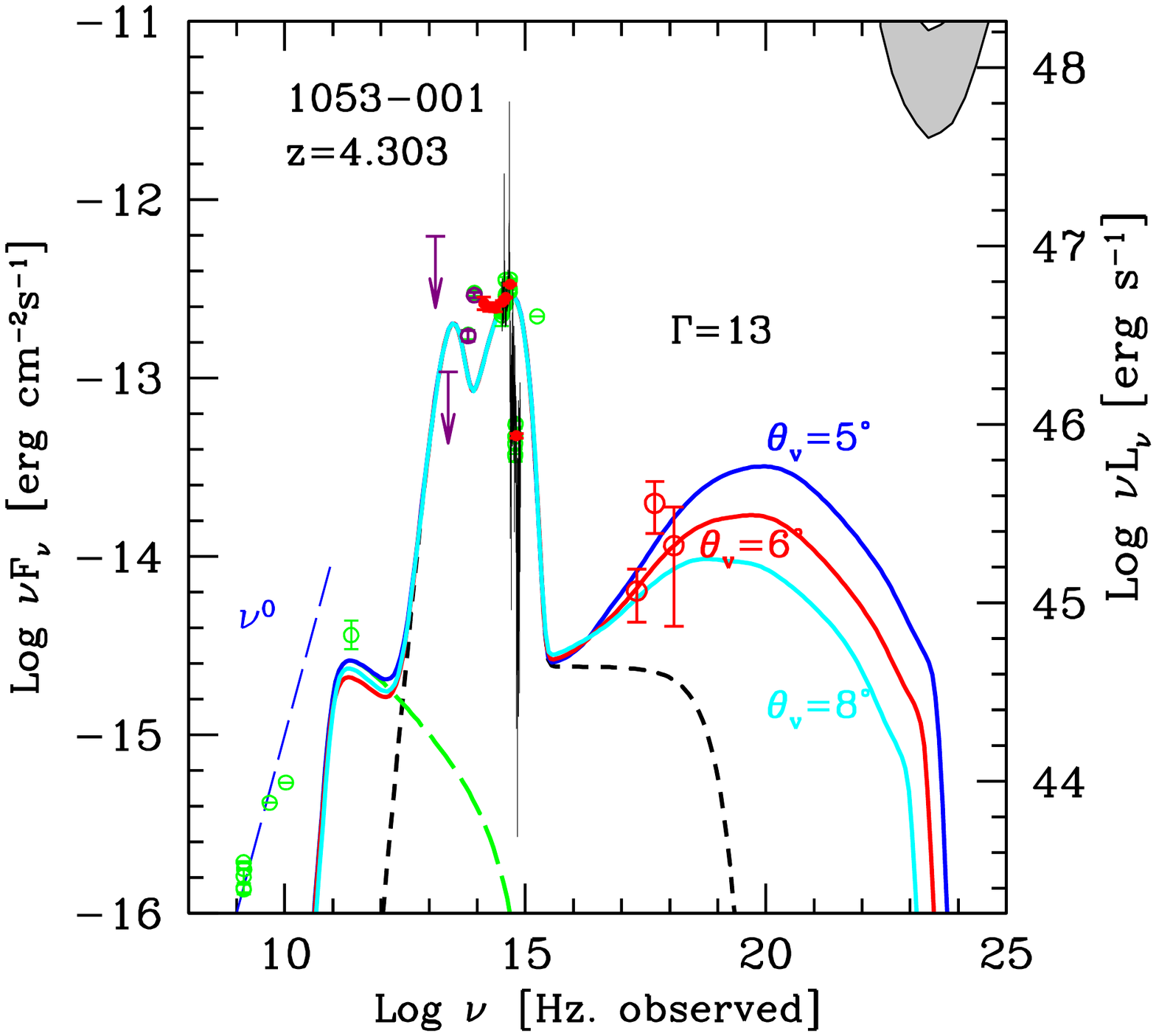} % ,width=12cm}
\vskip -2.5 cm
\caption{Broad band observed SEDs for sources  SDSS~J085111.59+142337.7 (left) and SDSS~J105320.42--001649.7 (right), 
         as examples of sources classified as a blazar and a non blazar, respectively.
         In all panels, the thin solid lines are SDSS spectra, purple points and upper limits 
         are from {\it allWISE}, red filled points are from GROND \citep{sbarrato13b}, 
         green points show archival data (Space Science Data Center). 
         Red empty data points are instead the new X--ray data from this work. 
         The models include a thermal emission from disk, jet and corona (dashed black line), 
         a synchrotron component (green lines) and an external Compton component at higher energies. 
         The different models in each panel differ mainly by their viewing angle, that mainly affect 
         the high energy section (different colours, as labelled). 
         The parameter of all solutions are detailed in Table \ref{tab:sed}.
} 
\label{fig:sed3}
\end{figure*}
%--------------------------------------------------

We  use the one--zone leptonic model by \citet{ghisellini09}.
In the original paper the reader can find a full discussion
of the features of the model.
We here define the model parameters. 
Note that the black hole mass $M$ and the accretion luminosity $L_{\rm d}$
are derived as discussed above.

\begin{itemize}

\item We assume that there is one region within the jet where most of
the luminosity is produced at a distance $R_{\rm diss}$ from the black hole.
We assume that the jet is conical  with fixed semi--aperture angle $\psi=0.1$.
Thus the size of the emitting (assumed spherical) region is $R=\psi R_{\rm diss}$.

\item The emitting region is magnetized with a tangled but homogeneous magnetic field $B$.

\item The emitting plasma moves with a velocity $\beta c$, corresponding to a 
bulk Lorenz factor 
$\Gamma$, at an angle $\theta_{\rm v}$ from our line of sight.

\item Throughout the region relativistic electrons are injected, with a power 
$P^\prime_{\rm e}$ as measured in the comoving frame. 
The injected distribution, assumed to be constant and homogeneous throughout the source, is:
\begin{equation}
Q(\gamma)  \, = \, Q_0\, { (\gamma/\gamma_{\rm b})^{-s_1} \over 1+
(\gamma/\gamma_{\rm b})^{-s_1+s_2} } \;\; \gamma_{\rm min} < \gamma < \gamma{\rm max}
\label{qgamma}
\end{equation}
where 
$s_1$, $s_2$, $\gamma_{\rm min}$, $\gamma_{\rm b}$, $\gamma_{\rm max}$ are the
slopes of the injected distribution of electrons (smoothly joining broken power law), 
and minimum, break and maximum random Lorentz factors. % of the SED.
The normalization $Q_0$ is set through 
\begin{equation}
P^\prime_{\rm e}= {4\pi \over 3} R^3 \int_{\gamma_{\rm min}}^{\gamma_{\rm max}} 
Q(\gamma) \gamma m_{\rm e}c^2 d\gamma 
\end{equation}
$\gamma_{\rm min}$ is set equal to one, while $\gamma_{\rm max}$, for $s_2>2$, is unimportant.
The slopes $s_1$ and $s_2$ are associated (by means of the continuity equation)
to the observed slopes before and after the two broad peaks of the SED.
The particle distribution is calculated at the dynamical time $R/c$.
This allows to neglect adiabatic losses, and allows us to use a constant (in time)
magnetic field and volume.

\item At a distance $R_{\rm BLR}$ there is the broad line region (BLR). 
We assume that it scales as:
\begin{equation}
R_{\rm BLR} \, =\,  10^{17} L_{\rm d, 45}^{1/2} \,\,\,{\rm cm}
\end{equation}
where $L_{\rm d, 45}$ is the disk luminosity in units of $10^{45}{\rm erg/cm^2/s}$
\item The molecular torus intercepts a fraction $f$ of $L_{\rm d}$ 
and re--emits it in the infrared. Its location is assumed to be:
\begin{equation}
R_{\rm torus}\, =\, 2\times 10^{18} L_{\rm d, 45}^{1/2} \,\,\,{\rm cm}
\end{equation}
\item We assume that the X--ray corona emits a luminosity $L_{\rm xc}$ in the form of a 
power law of energy index  $\alpha_X\sim 1$, ending in an exponential cut at 
$h \nu_{\rm c} \sim 150$ keV.
Usually we use 
$L_{\rm xc}(\nu)  \propto \nu^{-1}  \exp (- h\nu / 150\, {\rm keV} )$.

\end{itemize}
 
Of these parameters, 
$\psi$, $\theta_{\rm v}$, $L_{\rm xc}$, $\alpha_X$, $h \nu_{\rm c}$, $\gamma_{\rm min}$ 
are kept fixed (with rare exceptions).
The exact value of $\gamma_{\rm max}$ is unimportant (for $s_2>2$).
$L_{\rm d}$ is found through direct fitting or through the broad line luminosities,
and fixes $R_{\rm BLR}$, $L_{\rm torus}$, $R_{\rm torus}$, $\nu_{\rm ext}$.
The black hole mass $M$ is found through the disk fitting method
(when possible), or from the virial method.

We are left with 7 relevant free parameters:
$R_{\rm diss}$ (or equivalently, $R$), 
$B$, $\Gamma$, $P^\prime_{\rm e}$, $s_1$, $s_2$, $\gamma_{\rm b}$.
The observables used to constrain these parameters are:
\begin{itemize}
    \item the synchrotron and the inverse Compton luminosity ($L_{\rm S}$, $L_{\rm C}$);
    \item the synchrotron and the inverse Compton frequency peaks ($\nu_{\rm S}$, $\nu_{\rm C}$);
    \item the spectral indices pre and post peak ($\alpha_0$, $\alpha_1, $ that can be different for the synchrotron and IC peak, according to the relative importance of the SSC process and/or the importance of Klein--Nishina effects);
    \item the minimum variability timescales  ($t_{\rm var}$) when known.
\end{itemize}

The one zone model we are using never allows to account for the radio emission,
that  is due to a superposition of different emitting region at larger distances.

One relevant quantity that the model can return is the jet power, as the sum
of the power radiated and carried in the form of magnetic field, relativistic electrons
and protons (assumed to be cold):
All of these forms of powers can be written as:
\begin{equation}
P_{\rm i}\, =\, \pi (\psi R_{\rm diss})^2 U^\prime_{\rm i} \Gamma^2 c
\end{equation}
where $U^\prime_{\rm i}$ is the energy density of the $i$-component. 

\section{Results}
\label{sec:results}

Our main aim is to confirm or not the blazar nature of our candidates.
We are guided by SED modelling, that is in turn influenced by the shape and 
luminosity of the X--ray spectrum. 
Furthermore, we are interested to blazars whose black hole mass is larger than $10^9\, M_\odot$,
to find out the total number density of BH of large mass in radio--loud sources.

The left panel of Fig. \ref{fig:sed3} shows a typical example of a confirmed blazar. 
The X--ray flux is intense and hard, driving the derivation of a small viewing angle ($\theta_{\rm v}=3^\circ$), 
that combined with a Lorentz factor $\Gamma=13$ points toward a secure blazar classification. 
Even the strict definition that we follow is valid, in this case: the viewing angle is in fact smaller than the beaming angle ($\theta_{\rm v}\leq1/\Gamma$).
We also derive a black hole mass of $4\times 10^9 M_\odot$ and  
$L_{\rm d}=7 \% $ of the Eddington limit, by studying the thermal emission from the accretion disk. 

The right panel of Fig. \ref{fig:sed3}, instead, shows an example of a 
source whose jet is directed roughly in our direction, but does not fall in the strict definition of blazar.
The softer X--ray spectrum is the main responsible for a larger viewing angle classification. 
This is due to the large error bars on X--ray data: even if a large Lorentz factor of 13 is reasonable, 
the viewing angles consistent with multiwavelength data range between $5^\circ$ and $8^\circ$. 
SDSS~J105320.42--001649.7 is nevertheless an interesting source, having large mass and Eddington ratio 
($M\simeq1.5\times10^9M_\odot$ and $L_{\rm d}=27\%L_{\rm Edd}$), but cannot be taken into account in our upcoming discussions.

\subsection{Notes on single sources}

The SEDs of all the sources are reported in the Appendix, while in the following 
we review the results and subsequent classification of our sample, excluding the 
two sources described in the previous section and the one undetected in the X--rays. 
\\
{\bf SDSS~J081333.32+350810.8 ---} The BH mass is $4\times 10^9 M_\odot$ with 
$L_{\rm d}=12 \% L_{\rm Edd}$.
It can be fitted with $\theta_{\rm v}=5.2^\circ$, that is $\sim 1/\Gamma$, if $\Gamma=13$,
but also with  $\theta_{\rm v}=8^\circ$ (with the same $\Gamma$). 
Therefore it is likely a source viewed at the edge of our definition of blazar.
\\
%{\bf 085111.59+142337.7 ---} The BH mass is $4\times 10^9 M_\odot$ with 
%$L_{\rm d}=7 \% $ of Eddington. The strong and hard X--ray luminosity allows 
%to classify it as a blazar. 
%In fact, with $\Gamma=13$, the model with $\theta_{\rm v}=3^\circ$ fit the data well,
%while with $\theta_{\rm v}=6^\circ$ the model is below the X--ray data. 
%\\
{\bf SDSS~J103717.72+182303.0 ---} The BH mass is $10^9 M_\odot$ with 
$L_{\rm d}=20 \% $ of the Eddington luminosity. The data quality does not allow to definitely
classify this source as a blazar, because a range of $\theta_{\rm v}$
can fit the SED equally well, as long as $\theta_{\rm}<8^\circ$.
Therefore it is either a blazar or a source viewed at the border of our definition of
this class.
\\
%{\bf 105320.42--001649.7 ---} The BH mass is $1.5\times 10^9 M_\odot$ with 
%$L_{\rm d}=27 \% $ of Eddington. 
%The X--ray data have large error bars, and models with relatively large viewing angles
%can fit the spectrum. Therefore, conservatively, we classify this sources as a non--blazar. 
%\\
{\bf SDSS~J123142.17+381658.9 ---}  The BH mass is $M=7\times 10^8 M_\odot$ with
$L_{\rm d}=20\% L_{\rm Edd}$. 
We show three different models with the same $\Gamma=11$ but different $\theta_{\rm v}$.
According to our definition, to be classified as blazar, $\theta_{\rm v} \le 1/\Gamma$. 
Therefore, using $\Gamma=11$, the limiting angle is 5.2 degrees. 
We see that the model with $\theta_{\rm v}=8^\circ$ passes below the X--ray data,
that are instead well accounted for with smaller $\theta_{\rm v}$.
We thus classify this source as a blazar, but with a mass below $10^9 M_\odot$.
\\
{\bf SDSS~J123503.03--000331.7 ---} The BH mass is $M=2 \times 10^9 M_\odot$ with
$L_{\rm d}= 7\% $ Eddington. 
The X--ray spectrum is hard with a large luminosity, and is fitted well with $\Gamma=13$ 
and $\theta_{\rm v}=3^\circ$.
We classify this source as a blazar.
\\
{\bf SDSS~J124230.58+542257.3 ---} The BH mass is $M=3.5 \times 10^9 M_\odot$ with
$L_{\rm d}= 8\% $ Eddington. 
The X--ray spectrum is not well constrained, but data 
seem to exclude a very hard spectrum, therefore we classify the source as a non-blazar.
\\
{\bf SDSS~J141209.96+062406.8 ---}  The BH mass is  $M=6\times 10^8 M_\odot$
with $L_{\rm d}=63\%$ Eddington. 
We show two different models, one with $\Gamma=11$ and $\theta_{\rm v}=3^\circ$
and the other with $\Gamma=10$ and $\theta_{\rm v}=5^\circ$. 
The limiting $\theta_{\rm v}$ for $\Gamma=10$ is $5.7^\circ$.
We conclude that the source is indeed a blazar, but with a mass smaller than $10^9 M_\odot$.
\\
{\bf SDSS~J165913.23+210115.8 ---}  The BH mass is  $M=2.5\times 10^9 M_\odot$
with $L_{\rm d}=10\%L_{\rm Edd}$. 
The X--ray spectrum and luminosity clearly suggest the blazar nature of this source,
confirmed by our model with $\theta_{\rm v}=3^\circ$ and $\Gamma=13$.
\\
{\bf SDSS~J231448.71+020151.1 ---} The BH mass is  $M=1.5\times 10^9 M_\odot$
with $L_{\rm d}=15\%$ Eddington.  
The X--ray spectrum is hard and luminous: the best model prefer $\theta_{\rm v}< 1/\Gamma$,
but note that in this case we find a relatively small value of $\Gamma$ (=9),
so that the limiting angle becomes $6.3^\circ$.

\section{Double nature of high-$z$ blazar candidates}
\label{sec:double_jet}

\cite{cao17} have observed four $z>4$ blazar candidates 
with radio interferometry, in order to understand the radio jet orientation.
Interestingly, two sources that we classified as blazars in \cite{sbarrato15}, 
when observed at high resolution, appear as misaligned jetted sources. 
X--ray emission clearly shows a strong and hard spectrum, that is a clear signature of 
a relativistic jet aligned to our line--of--sight. 
Radio emission instead shows features of misaligned relativistic jets. 
This puzzling discrepancy highlights some difficulties in interpreting the jet orientation at high redshift. 
We consider two possibilities to interpret these observational features:
\begin{itemize}
\item[-] we are observing blazars whose jet is bent, or underwent a change in direction, with a high--energy emitting region aligned to our line--of--sight, differently from the extended radio emission orientation; 
\item[-] the relativistic jet structure and emitting features are intrinsically different at low and high redshift, leading to different observational signatures.
\end{itemize}

We cannot in fact exclude that jets may have different emitting features at high redshift: no wide multiwavelength population studies are possible yet, with few hundreds sources known, of which only a fraction has multiwavelength jet studies. 
Up to $z\sim3$ the features of jetted sources do not appear to be drastically different, though. 
In other words, the jet emission is produced by a region moving along the jet with bulk Lorentz factor in the range $\Gamma\sim10-15$, with a beaming angle of approximately $1/\Gamma$.
If the jet is observed at small viewing angles (or even inside the beaming cone), the AGN broad band flux will likely be dominated by the jet emission, and specifically the X--ray flux will be intense and with a hard spectrum, while the radio emission will be flattened, bright and compact on VLBI scales. 
If the viewing angle is larger (e.g.\ $>10^\circ$), the X--ray spectrum will be significantly dimmer and softer, easily going undetected even at smaller redshift, up to the point where the X--ray corona emission will dominate.
The sources we classified as blazars both in this work and in \cite{sbarrato15} show strong and hard X--ray fluxes, that together with the broad--band SED modeling have led us to assume strong beaming and small viewing angles.  
VLBI observations by \cite{cao17}, instead, show for some sources larger viewing angles, up to $20^\circ$, clearly not consistent with bulk Lorentz factors $\Gamma\sim10$.
We cannot exclude that jets are less extremely relativistic at higher redshift, with smaller bulk Lorentz factors. 
This would widen the beaming cone of these blazar candidates, allowing the X--ray flux to stay brighter up to much larger viewing angles, but simultaneously losing hardness in the X--ray spectrum.
Again, a large viewing angle would not be consistent with the X--ray emission that we observe for these sources. 

The other, maybe easier, possibility is that the physical features of AGN jets are still coherent with the low redshift population, but the relativistic jets are bent more easily. 
AGN with bent jets have already been observed in the local Universe.
The bending might be due to changes in the propagation medium densities, or in the jet base orientation, i.e.\ possible changes in the black hole spin orientation or in the black hole--accretion disk relative orientation. 
If, as one would expect in an extremely fast formation and evolution scenario, the AGN environment is more chaotic than in a lower redshift counterpart, the jet will more likely be bent by one of these conditions.
Under this assumption, the X--ray emitting region, closer to the central SMBH, would be aligned to our line--of--sight.
Moving farther away from the nuclear region, instead, the jet might be bent and the radio emission would come from a region that is therefore directed away from our line--of--sight, or at least at larger viewing angles. 
This geometric configuration would explain why the broad band SED, with particular emphasis over the high--energy part, is better described by an aligned relativistic jet, while the high-resolution radio interferometry hints toward a jet misaligned up to $20^\circ$.

The available data do not allow us to disentangle between these options. 
More extensive classifications of jetted AGN candidates at $z>4$ performed independently with radio and X--ray based methods are paramount, if only to establish the statistics of the inconsistencies. 
X--ray data are already on the edge of {\it Swift}/XRT sensitivity for many of the blazar candidates we are considering: more powerful and upcoming facilities will certainly be helpful in deepening our understanding of the high--energy component.

\section{Early evolution of jetted AGN}
\label{sec:early_jet}

%--------------------------------------------------
\begin{figure} 
\hskip -0.5 cm
\includegraphics[width=1.15\hsize]{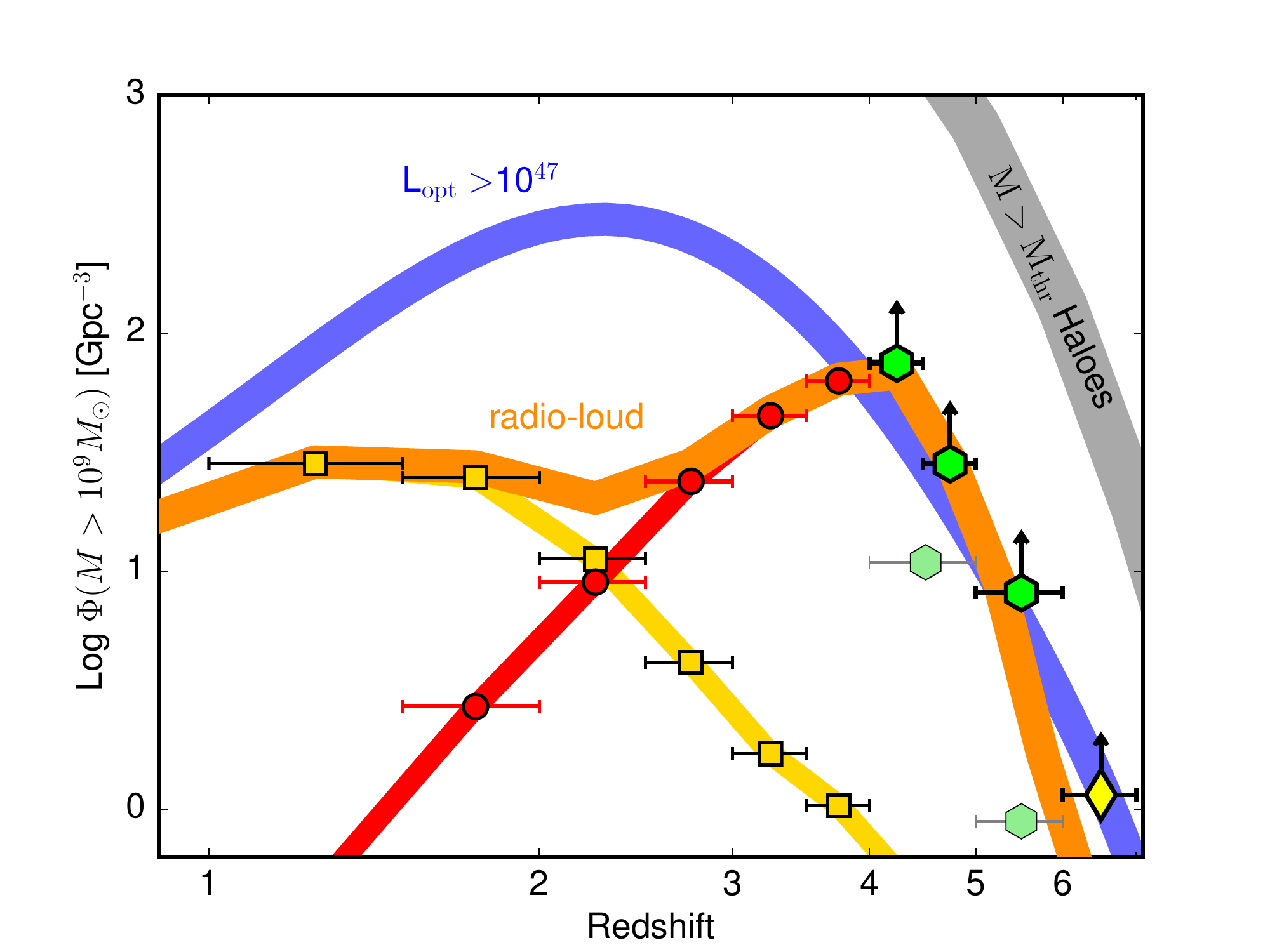} % ,width=12cm}
\caption{Comoving number density of $M_{\rm}\geq10^9M_\odot$ SMBHs hosted in jetted 
         (orange line) AGN and in all AGN (blue line).
         The non--jetted curve is obtained from \cite{shen20}.
         The radio--loud sample is instead composed by the {\it Fermi}/LAT 
         (yellow squares and line) and {\it Swift}/BAT (red circles and line) 
         all--sky catalogs.
         The bright green hexagons with thick black contours are lower limits 
         to the $z>4$ comoving number density of jetted sources 
         derived by known blazars included in the spectroscopic SDSS+FIRST footprint, 
         updated at the results in this work. 
         The less contrasted hexagons show instead the state of the art earlier 
         than systematic searches of high--$z$ blazars begun.
         Finally, the yellow diamond is the lower limit obtained 
         by the detection of PSO~J0309+27 \citep{belladitta20}.
         The data points obtained by systematic search and classification of 
         blazar candidates are lower limits, being the search still ongoing.
         } 
\label{fig:phi}
\end{figure}
%--------------------------------------------------

The large majority of discovered quasars at $z>4$ have their black holes with masses 
larger than $10^9M_\odot$.
This includes our sources, but SDSS~J123142.17+381658.9 and SDSS~J141209.96+062406.8, that have masses of about $7$ and $6\times10^8M_\odot$, respectively\footnote{
The uncertainty on the accretion disk emission based masses is of the order of a factor 2, so the sources with $<10^9M_\odot$ might still be consistent with $10^9$ solar mass SMBHs. Nevertheless, we prefer to be conservative and not include them in the following discussion.}.  
Note that this is not an instrinsic property of the high--redshift quasar population, 
but it is an effect of the observing bias that follows telescopes sensitivity limits: 
we are only observing the most luminous, hence the most powerful and likely more 
massive of the quasar population. We are mapping only the tip of the iceberg. 

It is nonetheless interesting to investigate the objects in this mass range: following \cite{ghisellini10,ghisellini13}, \cite{sbarrato15}, \cite{sbarrato21} we aim at focusing on the distribution of such sources. 
We extracted the comoving number density of quasars with masses $>10^9M_\odot$ from the luminosity function of quasars built by \cite{shen20}. 
We aimed at including theoretical luminosity functions, but they usually follow observations, e.g.\ \cite{yung21} reproduces the distribution obtained by \cite{shen20}, and it would thus not add any information. 

Since we do not have black hole mass estimates for the whole sample by \cite{shen20}, we assume that all sources with a bolometric luminosity larger than the Eddington luminosity for a $10^9M_\odot$ SMBH (i.e.\ $L_{\rm opt}>10^{47}$erg/s have $M>10^9M_\odot$) indeed host only $>10^9M_\odot$ SMBH. 
Therefore, we assumed that all sources with $L_{\rm opt}>10^{47}$erg/s have $M>10^9M_\odot$. 
Clearly we will be missing billion solar masses sources with smaller luminosities (and hence smaller Eddington ratios), but the object density will not be contaminated by black holes with smaller masses. 
The number density that can be extracted from this sub--sample is thus a lower limit on the total $10^9M_\odot$ population, but not far in luminosity from the quasars currently observed at high redshift. 
This approach allows to study the whole quasar population: no distinction between jetted and non--jetted sources is made in the work by \citet{shen20}.

In order to study the density of only jetted sources, instead, we chose to focus on blazars, in order to follow the same approach at all redshifts. 
Defining blazars only as those jetted AGN with their viewing angles ($\theta_{\rm v}$) smaller than their beaming angles ($1/\Gamma$), it can be inferred the presence of $2\Gamma^2$ analogous sources randomly oriented in the sky. 
Since typically for these powerful sources $\Gamma\sim10-15$, hundreds of jetted AGN can be inferred by the detection of even few blazars. 
Thus, instead of considering a general sample of  jetted sources, we only focus on blazars: up to $z \sim3.5$, all--sky high--energy catalogs are available, while at higher redshifts one must consider individually classified sources. 
For the lower redshift part, we again followed \cite{ghisellini10} and used the same {\it Fermi}/LAT and {\it Swift}/BAT all--sky catalog releases, while the $z\geq4$ part is composed by our systematic classification. 

Figure \ref{fig:phi} shows the comoving number density of these samples, immediately
highlighting that the distribution of jetted and non--jetted $M_{\rm BH}>10^9M_\odot$ 
quasars are strikingly different. 
The total population, dominated by the radio--quiet and silent sources, shows a prominent 
peak at $z\sim2.5$, as expected. 
On the other hand, jetted sources, as traced by blazars, appear to have a different
density distribution across cosmic history \citep[maybe affected by the combination of two different all--sky samples, ][]{sbarrato15} and in particular a much larger 
comoving number density at $z\sim4$ than in the closer Universe. 
This was already evident in previous versions of this analysis, but the striking 
news with the updated classification is that jetted sources are as numerous as the 
$L_{\rm opt}>10^{47}$~erg/s population\footnote{
This result is extreme since the Lorentz factors of our sources are, as expected, between 10 and 15. 
For a more stringent consistency with the low redshift jetted--to--non--jetted ratio, smaller $\Gamma$ values are needed, or not limiting the sample to the largest masses \citep{diana22}.
}.
If these samples were complete, we should conclude that most $>10^9M_\odot$ black holes at $z>4$ are hosted in jetted quasars, in spite of their radio--loudness. 
Note that $10^{10}M_\odot$ z>5 quasars have already proved to host relativistic jets even when they are radio--quiet or when they completely lack standard jet signatures \citep{sbarrato21VLA}. At high redshifts, the same behaviour might extend toward smaller masses (but still above a billion $M_\odot$), even only for a substantial fraction of them. 

We cannot completely exclude that the issue resides in underestimating non--jetted sources: it has already been observed that high redshift quasars are indeed over-obscured with respect to their ``local" counterparts \citep[e.g.\ ][]{mortlock11,fan20}. 
The issue has already been investigated, and it has been suggested that a ``dust bubble" (or more in general an over-obscuring dusty region) surrounding the accreting nucleus might correct the observed discrepancy between the number of blazars observed and the expected parent population that is currently lacking in radio catalogs \citep{fabian99, ghisellini16}. 
That hypothesis suggested that accreting supermassive black holes (specifically at very high redshift) are surrounded by an excess of obscuring dust, isotropically distributed. 
When the accretion luminosity of the nascent quasar reaches a threshold luminosity, its radiation pressure and sublimation power manage to get rid of the accumulated dust. 
On the other hand, while below the threshold the accreting black hole stays obscured, 
\citet{ghisellini16} suggested that during the over-obscuration phase the jet would be able to pierce through the bubble. In this case, only very aligned jetted sources are visible in the optical wavelength range.
The same mechanism might uniformly apply to non--jetted sources. In that case, quasar luminosity functions would all need a more pronounced correction in order to estimate the absorbed (and thus not observed) quasars. 
A much too wild speculation would be needed to estimate whether this might be enough to change the profile of distributions such as the comoving number density of non--jetted sources. 
Nonetheless, even with the contribution of over-obscuring bubbles, the jetted population would likely be a larger fraction of the overall quasar one, with respect to lower redshift distributions. 

\subsection{Jets and fast mass accretion rates}

The large occurrence of relativistic jets in extremely massive quasars at high redshift is compelling, and it reopens the question of how jets affect the evolution of early massive black holes. 
If the most massive black holes actually prefer to form in jetted sources, jets might play a crucial role in their early formation or be the smoking gun for particularly favorable fast accretion conditions. 
A simple toy model to interpret the possible jet role in black holes fast accretion has already been explored in the last few years \cite[e.g.\ ][]{ghisellini13,mazzucchelli17,sbarrato21}.
The core concept relies on the re--distribution of the gravitational energy released during the accretion process. 
Normally, the accretion disk emission is thought to be produced with a radiative efficiency coincident with the energy release fraction from the accretion structure. 
If instead the relativistic jet launch and acceleration processes exploit part of the gravitational energy release, 
not all of that is radiated away, allowing a faster accretion for the same observed luminosity. 
This would easily speed up the most massive black hole formation, but it is still not clear how it works. 

Another option might reside in the jet interaction with the environment: what if the jet itself is able to trigger the infall of galactic matter in the vicinity of the black holes?
It is already well known that jet activity influences the host galaxy environment, likely being a key actor in its evolution. 
Observational signatures of both positive and negative feedback have been observed in local active galaxies \cite[see e.g.\ ][]{fabian12,silk98,harrison18,yuan18,cresci18}. 
The possibility that a relativistic jet is responsible for the infall of matter towards the nucleus, after affecting star formation activity, is thus not unlikely. 
Under this assumption, a jet would be indirectly responsible for fast accretion, but the timing of jet and super--critical accretion phases would also be critical in speeding up the process. 
Structured extended sources or at least signs of jet extension in the nucleus surroundings would be the smoking gun for this hypothesis.

Understanding the mutual influence of accretion and jet, and the role of the latter in fast black hole evolution are some of the upcoming challenges of AGN physics. 
Higher resolution studies of the host galaxy morphology and star formation phase, other than of the extended jet structure itself will be crucial to understand the relative age of the relativistic jet and the supermassive black hole. 
Detailed studies of the regime at which high--z jetted quasars are accreting will also be instrumental, in combination with ionisation studies of the nuclear AGN region. 
This might help in disentangling the cause--effect conundrum: is the jet responsible for triggering a fast accretion phase, or is the black hole that forces the formation of a relativistic jet by pushing a close--to-- or super--critical accretion regime? 
Upcoming large facilities leading to much deeper optical and IR photometry and spectroscopy (ELT, JWST, VRO) will allow huge steps towards our understanding of the first massive black holes evolving at the Universe dawn.

\section{Conclusions}
\label{sec:conclusion}

We observed in X--rays, collected archival data and classified eleven extremely radio--loud $z>4$ blazar candidates from the SDSS+FIRST survey. 
After modeling their broad band SEDs, we were able to classify them as blazars or slightly misaligned jetted AGN, in order to include them in our study of the jetted quasar population at $z>4$. 
One source was not detected by {\it Swift}/XRT, five were classified as blazars, while two have an uncertain viewing angle and three are observed outside their beaming cone.
Two blazars host a black hole with less than a billion solar masses.

Some sources that we previously classified as part of this project were also observed with VLBI, and are classified differently if VLBI observations are considered. 
While high--energy emission points toward small viewing angles, in some cases high--resolution radio data suggest completely misaligned relativistic jets.
We discuss that the chaotic formation and fast accretion of these systems, as might occur at such high $z$, might be responsible for a more frequent bending of AGN jets. 
With current facilities, we can only enlarge the statistics in order to assess the occurrence of such discrepancies.

We finally focused on the jetted--to--non--jetted ratio across a wide span of redshifts, finding that at $z>4$ jetted sources outnumber the non--jetted ones among $>10^9M_\odot$ active black holes. 
This proves that jets must have a prominent role in the fast formation of the first supermassive black holes in the Universe. 
We speculate that jets directly affect accretion of matter onto the black hole, by facilitating the dispersion of gravitational energy released during the accretion process itself, allowing for a faster mass increase at the observed disk luminosity. 
We cannot exclude that instead it plays a role only in the infall of matter toward the galactic nucleus by affecting star formation on larger scales, but the sources we are currently studying show both active relativistic jets and fast accreting central black holes acting simultaneously, pointing towards a more intimate link between the two structures. 

\begin{acknowledgements}

We thank the referee for their useful and insightful comments that helped us in improving the paper.
We acknowledge financial support from ASI grant I/004/11/5. 
We made use of data supplied by the UK {\it Swift} Science Data
Centre at the University of Leicester.
Part of this research is based on archival data, software and
online services provided by the Space Science Data Center (SSDC), 
on the NASA/IPAC Extragalactic Database (NED)
and of the data products from the {\it Wide-field Infrared Survey
Explorer}, which are operated by the Jet Propulsion Laboratory,
Caltech, funded by the National Aeronautics and Space Administration.

\end{acknowledgements}


\bibliographystyle{aa} % style aa.bst
\bibliography{BIB} % your references Yourfile.bib

%%\begin{appendix}
%%\section{Broad--band SEDs}
\section*{Appendix A: Broad--band SEDs}

We collect in this section all broad--band SEDs of the sources from our sample that have been 
detected by {\it Swift}/XRT, thus excluding SDSS~J030437.21+004653.5.
All figures show in labels the crucial parameters useful to classify them as blazars or not, 
i.e.\ viewing angle and Lorentz factor.

%--------------------------------------------------
\begin{figure*}
\vskip -2.7 cm
\hskip -0.1 cm
%\vskip -2.5 cm
%\hskip -0.25 cm
\includegraphics[width=9.2cm]{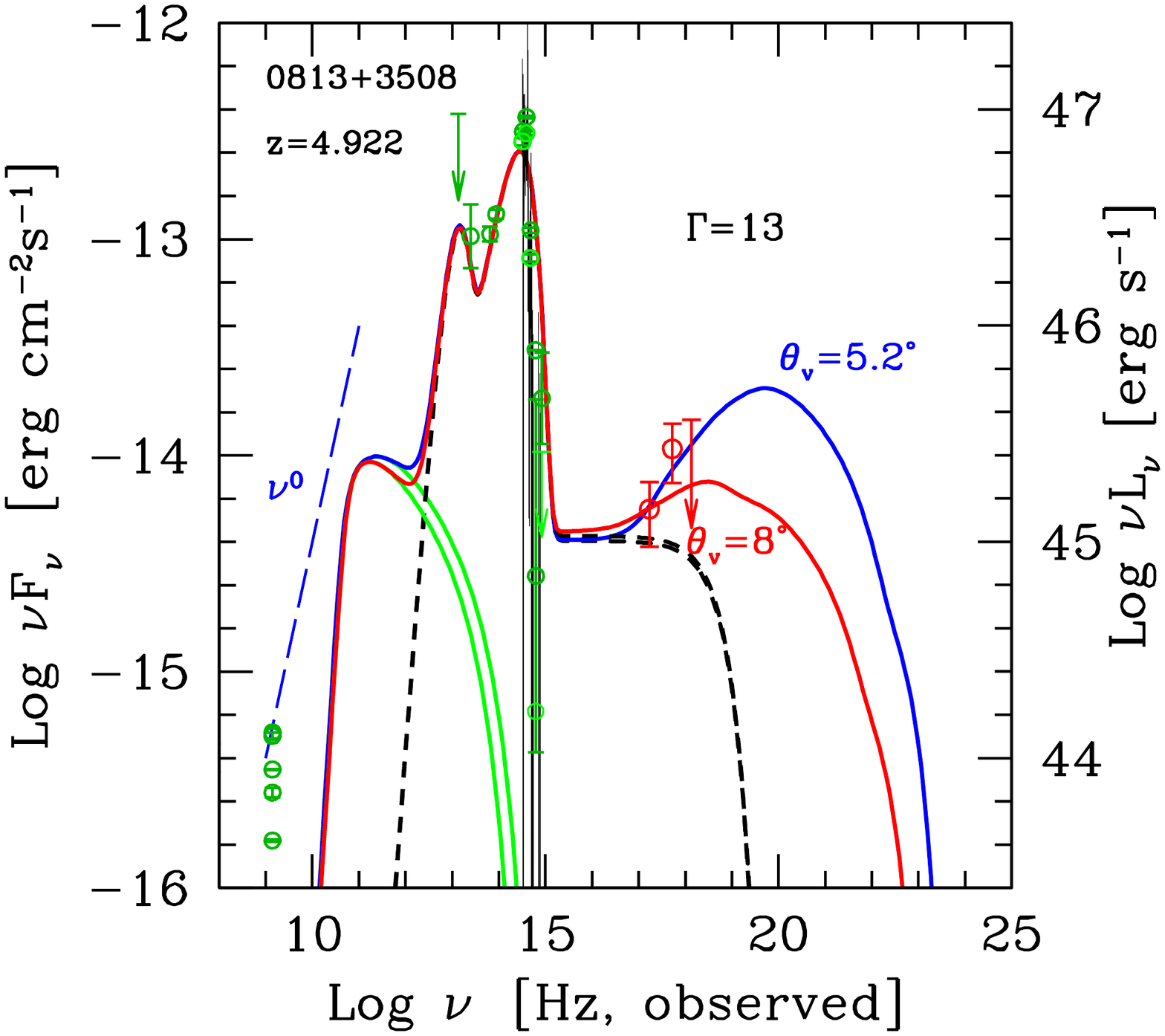} % ,width=12cm}
%\vskip -4.5 cm
\hskip -0.25 cm
\includegraphics[width=9.2cm]{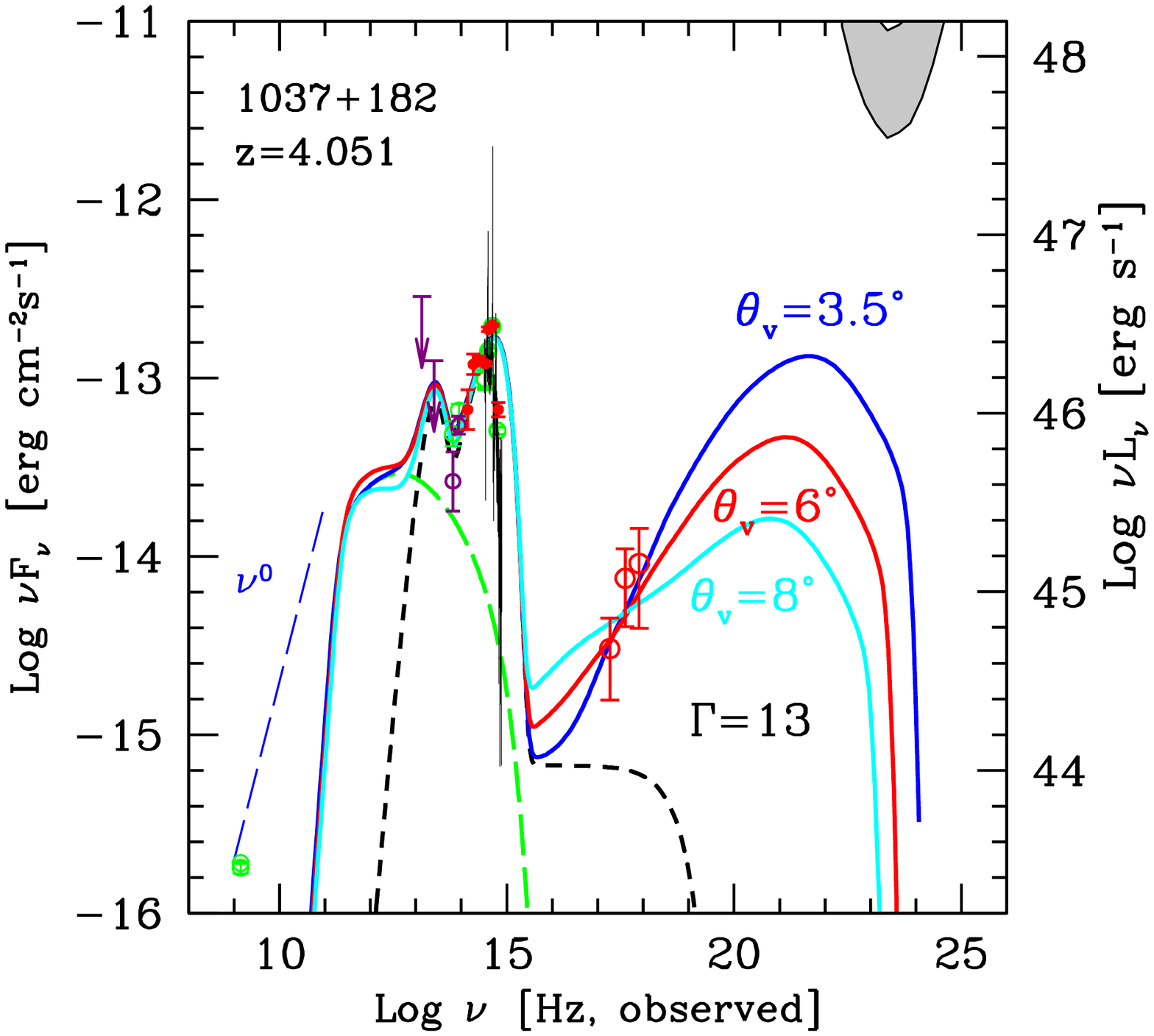} % ,width=12cm}
\vskip -5 cm
\hskip -0.1 cm
\includegraphics[width=9.2cm]{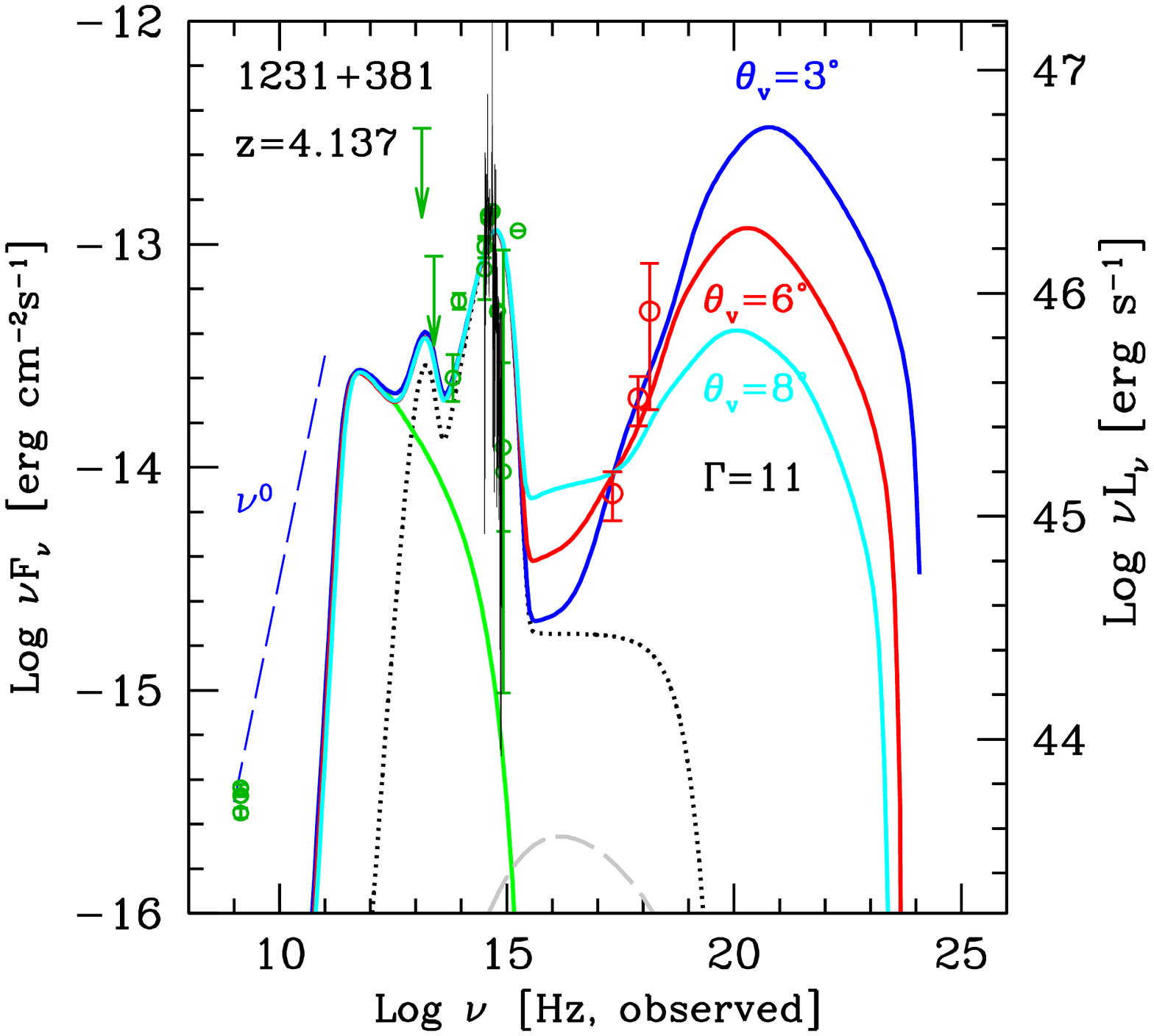} % ,width=12cm}
\hskip -0.25 cm
\includegraphics[width=9.2cm]{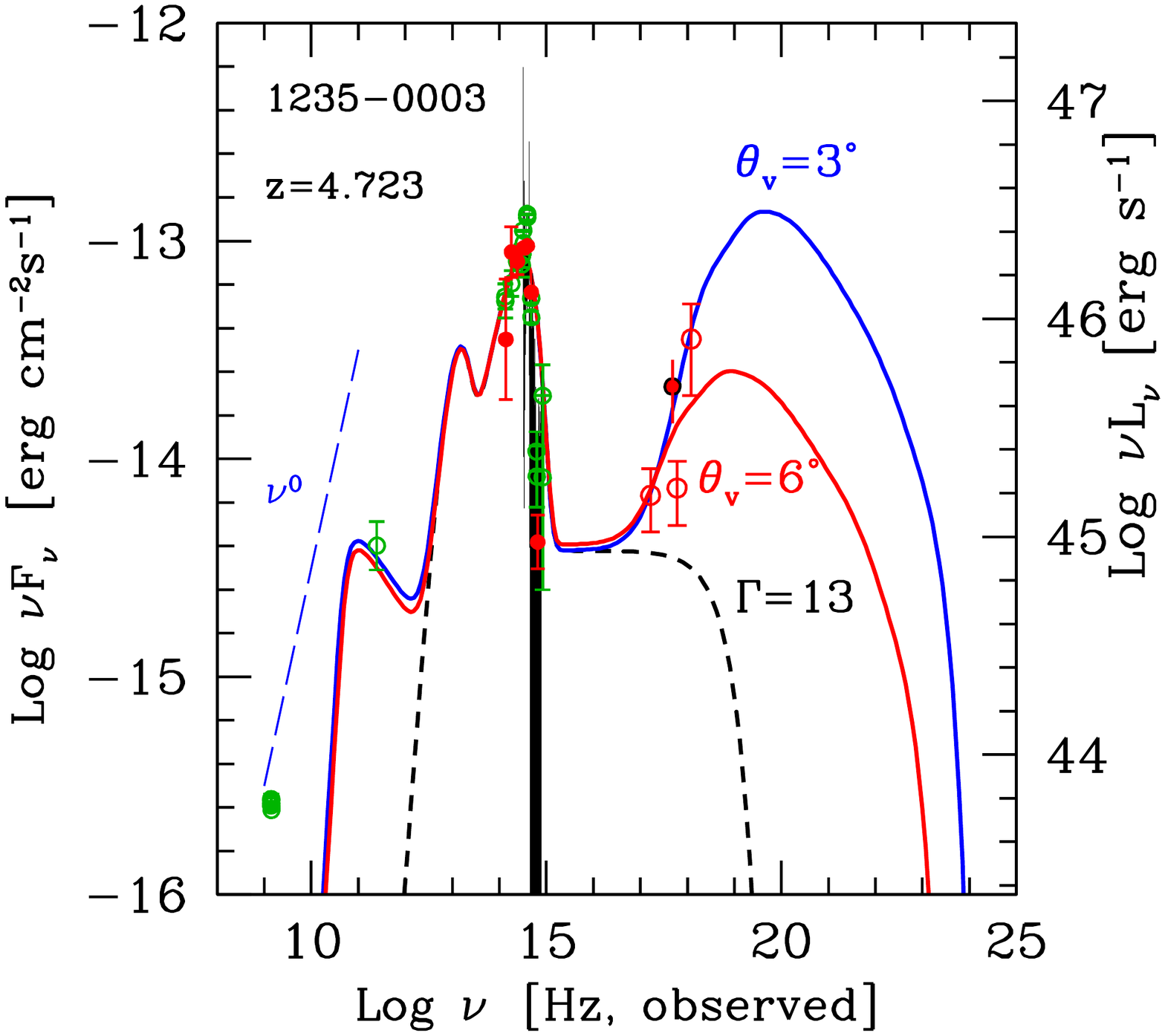} % ,width=12cm}
\vskip -2.5 cm
\caption{Broad band observed SEDs of the candidates
         SDSS~J081333.32+350810.8, SDSS~J103717.72+182303.0, SDSS~J123142.17+381658.9 and SDSS~J123503.03--000331.7,
         as labelled in shortened versions. Color, points and line coding as in Figure \ref{fig:sed3}.
        } 
\label{fig:sed1}
\end{figure*}
%--------------------------------------------------
%--------------------------------------------------
\begin{figure*} 
\vskip -2.7 cm
\hskip -0.1 cm
\includegraphics[width=9.2cm]{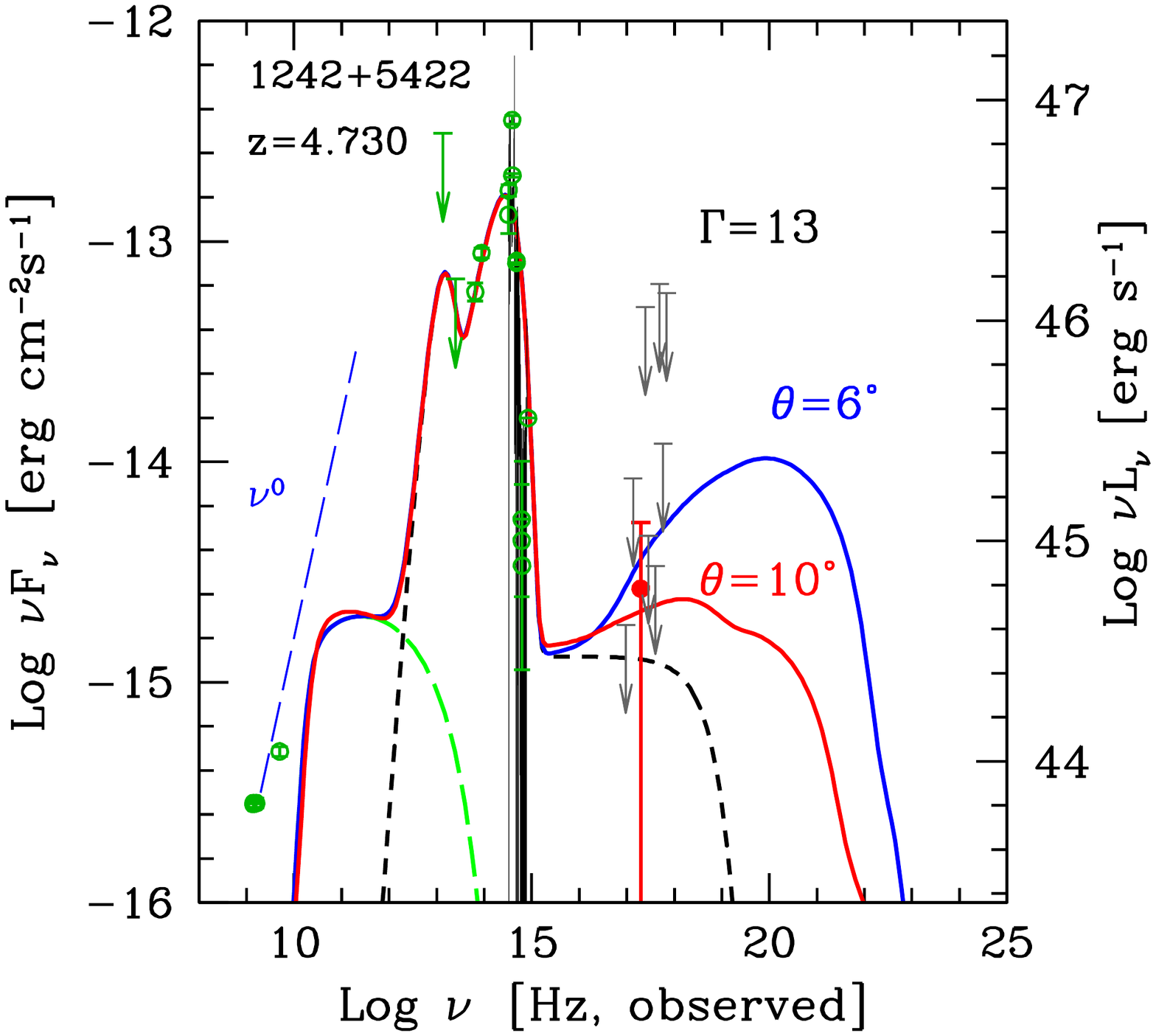} % ,width=12cm}
\hskip -0.25 cm
\includegraphics[width=9.2cm]{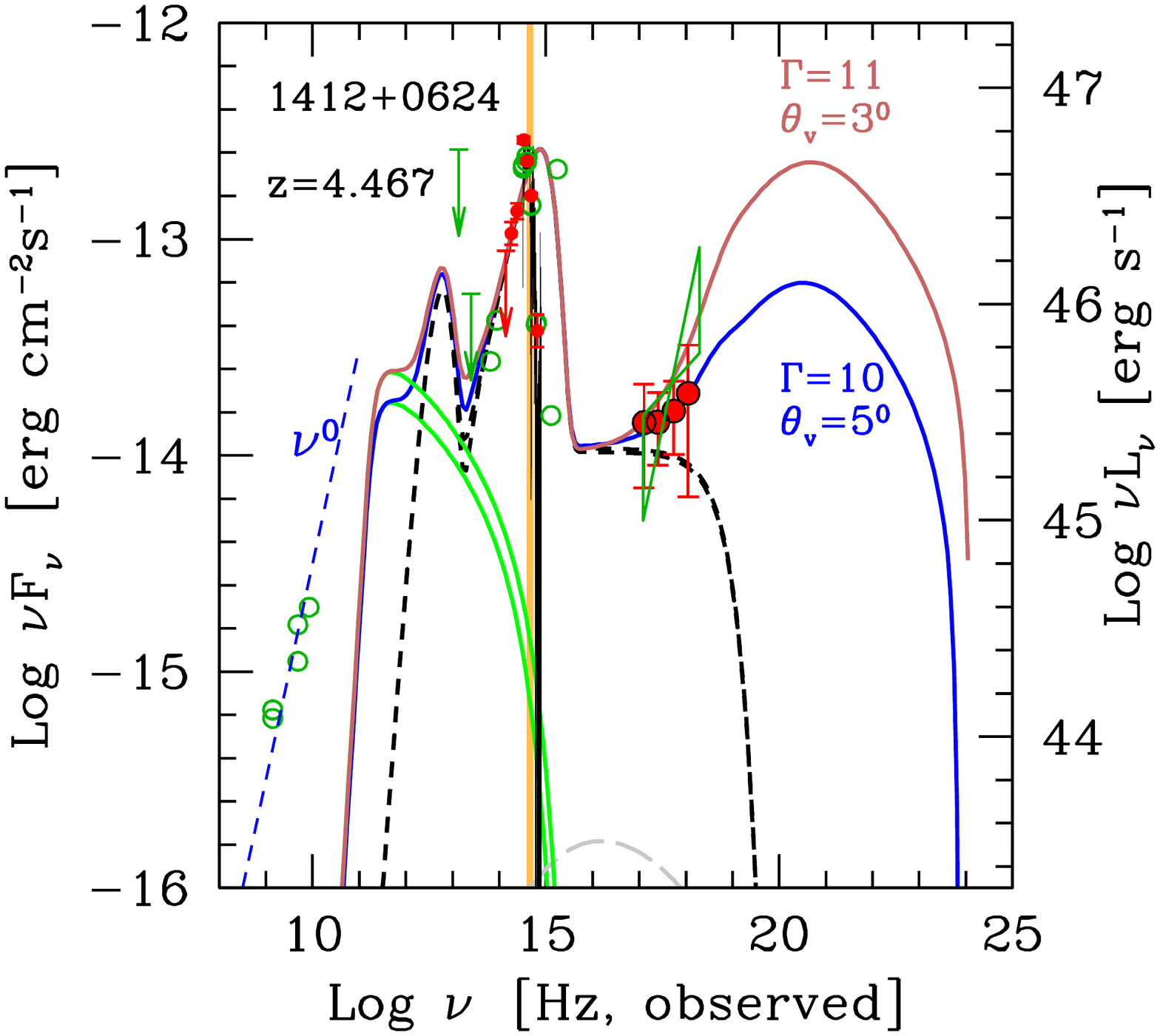} % ,width=12cm}
\vskip -5 cm
\hskip -0.1 cm
\includegraphics[width=9.2cm]{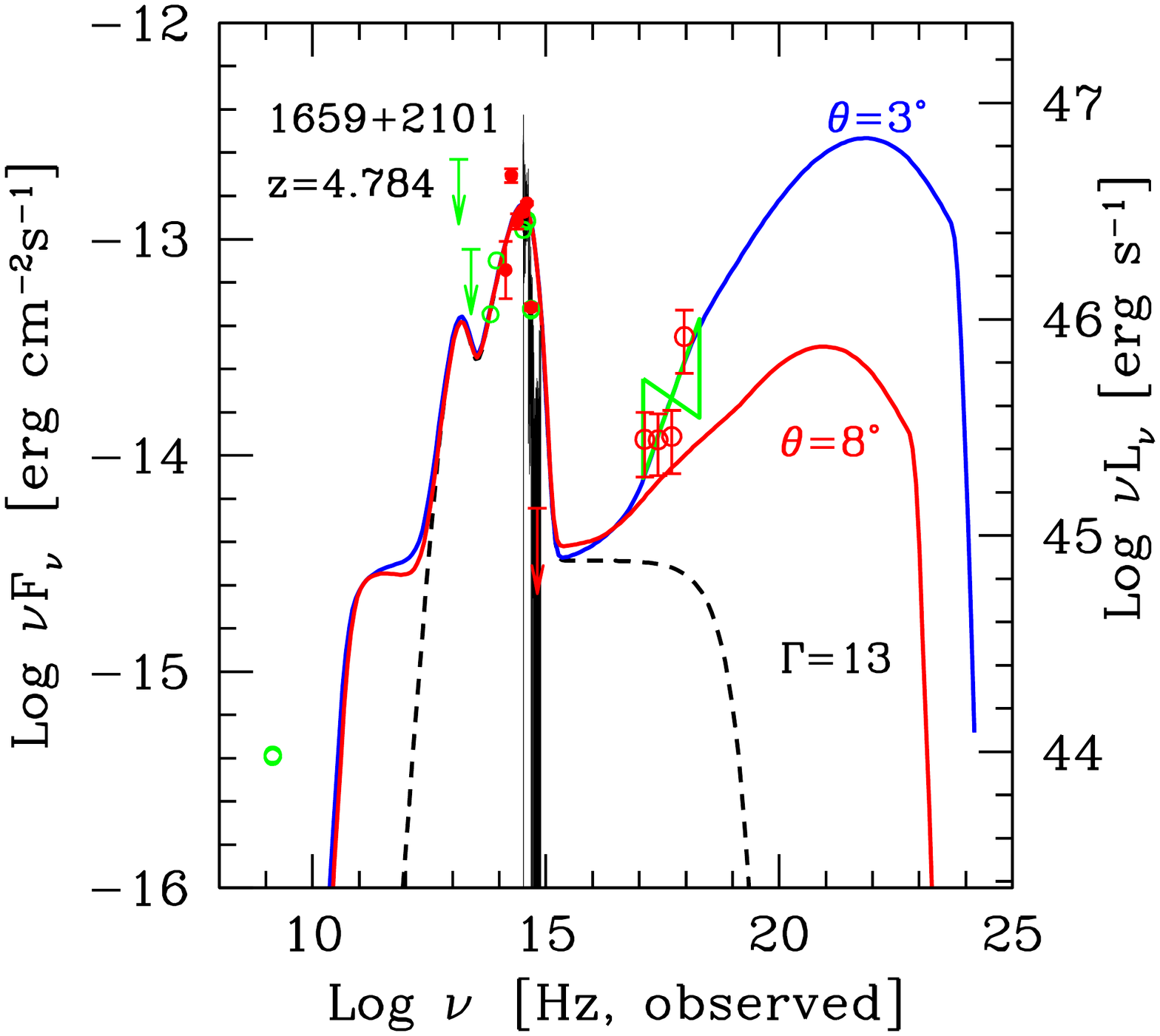} % ,width=12cm}
\hskip -0.25 cm
\includegraphics[width=9.2cm]{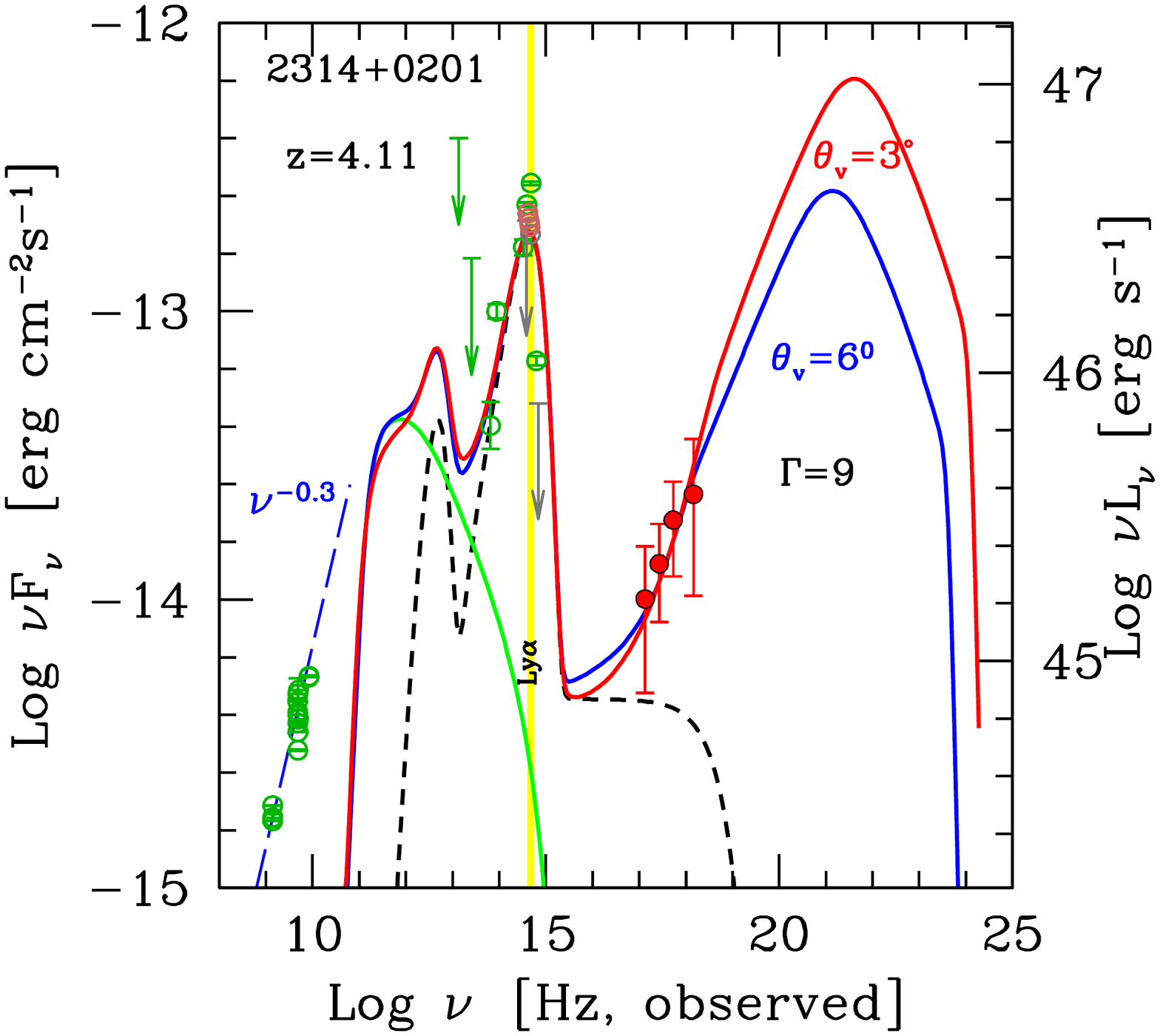} % ,width=12cm}
\vskip -2.5 cm
\caption{Broad band observed SEDs of the blazar candidates 
         SDSS~J124230.58+542257.3, 
         SDSS~J141209.96+062406.8, SDSS~J165913.23+210115.8 and SDSS~J231448.71+020151.1. 
         Color coding is the same as in Figure \ref{fig:sed3}
        } 
\label{fig:sed2}
\end{figure*}
%--------------------------------------------------
%%--------------------------------------------------
%\begin{figure*} 
%\vskip -2.5 cm
%\hskip -0.25 cm
%\includegraphics[width=9.4cm]{1659_21f.pdf} % ,width=12cm}
%\hskip -0.25 cm
%\includegraphics[width=9.4cm]{2314_020f.pdf} % ,width=12cm}
%\vskip -2.5 cm
%\caption{Broad band SEDs of the candidates SDSS~J165913.23+210115.8 and SDSS~J231448.71+020151.1, 
%         as labelled in shortened versions. Color coding is the same as in Figure \ref{fig:sed3}.
%        } 
%\label{fig:sed1}
%\end{figure*}
%%--------------------------------------------------

%%\end{appendix}

\end{document}